\documentclass[status=final,submission,copyright,creativecommons]{eptcs}

\usepackage[svgnames]{xcolor}
\usepackage[cmex10]{amsmath}
\usepackage{graphicx}
\usepackage[para]{footmisc}
\usepackage{booktabs}
\usepackage{tabularx}
\usepackage{multicol}
\usepackage{cleveref}
\Crefname{section}{Sect.}{Sects.}
\Crefname{equation}{Formula}{Formulae}
\Crefname{lstlisting}{Listing}{Listings}

\usepackage{xspace}
\newcommand{\cf}{cf.\xspace}
\newcommand{\ie}{that is,\xspace}
\newcommand{\eg}{for example,\xspace}
\newcommand{\wrt}{with respect to\xspace}
\newcommand{\ies}{i.e.,\xspace}
\newcommand{\egs}{e.g.\xspace}
\newcommand{\Eg}{For example,\xspace}

\newcommand{\etc}{etc.\xspace}

\usepackage{tikz}
\usetikzlibrary{automata,calc,arrows.meta}

\tikzset{
  stlab/.style={circle,fill=cyan!30}
}

\newcommand{\rfct}[1][f]{\mathsf{#1}}
\newcommand{\inact}[1][f]{0^{\mathit{#1}}}
\newcommand{\activ}[1][f]{\mathit{#1}}
\newcommand{\mitig}[1][f]{\overline{\mathit{#1}}}
\newcommand{\mishp}[1][f]{\underline{\mathit{#1}}}

\newcommand{\yapkeyw}[1]{\textcolor{blue}{\texttt{#1}}}

\newcommand{\yapag}[1]{\textbf{{\upshape}#1}}
\newcommand{\yapact}[1]{\textsf{{\upshape}#1}}
\newcommand{\yapcmd}[1]{\textcolor{red}{{\upshape}\textsf{#1}}}
\newcommand{\yaptplparam}[1]{\texttt{<\%YAP\##1>}}
\newcommand{\yapitem}[1]{\textbf{{\upshape}#1}}

\usepackage{acro}
\acsetup{
format-include-endings = true,
group-citation = true}
\DeclareAcronym{CE}{short = CE, long = critical event}
\DeclareAcronym{CPN}{short = CPN, long = coloured Petri net}
\DeclareAcronym{DTMC}{short = DTMC, long = discrete-time Markov chain}
\DeclareAcronym{DSL}{short = DSL, long = domain-specific language}
\DeclareAcronym{HARA}{short = HARA, long = hazard analysis and risk assessment}
\DeclareAcronym{HazOp}{short = HazOp, long = hazard operability studies}
\DeclareAcronym{HRC}{short = HRC, long = human-robot collaboration}
\DeclareAcronym{LTS}{short = LTS, long = labelled transition system}
\DeclareAcronym{MDP}{short = MDP, long = Markov decision process, long-plural=es}
\DeclareAcronym{MCSeq}{short = MCSeq, long = minimal cut sequence}
\DeclareAcronym{PCTL}{short = PCTL, long = probabilistic computation tree logic}
\DeclareAcronym{pGCL}{short = pGCL, long = probabilistic guarded command language}
\DeclareAcronym{PRA}{short = PRA, long = probabilistic risk assessment}
\DeclareAcronym{ROS}{short = ROS, long = robot operating system,
  long-post = {\footnote{See \url{https://www.ros.org}.}}}
\DeclareAcronym{STPA}{short = STPA, long = System-Theoretic Process Analysis}

\newcommand{\Yap}{\textsc{Yap}\xspace}
\newcommand{\Prism}{\textsc{PRISM}\xspace}

\usepackage{listings}
\lstdefinelanguage{prism}{
  morekeywords={
    A, bool, clock, const,
    ctmc, C, double, dtmc, E, endinit, endinvariant, endmodule, endrewards, endsystem,
    false, formula, filter, func, F, global, G, init, invariant, I, int, label, max, mdp,
    min, module, X, nondeterministic, Pmax, Pmin, P, probabilistic, prob, pta, rate,
    rewards, Rmax, Rmin, R, S, stochastic, system, true, U, W
  },
  morecomment=[l]{//}}
\lstdefinestyle{prism}{
  belowskip=0em,
  frame=tb,
  framesep=3pt,
  mathescape=true,
  breaklines=true,
  basicstyle=\sffamily\footnotesize,
  keywordstyle=\bfseries,
  identifierstyle=\slshape,
  commentstyle=\color{gray},
  emphstyle=\bfseries,
  tabsize=4,
  numbers=none,
  numberstyle=\tiny,
  stepnumber=1,
  numbersep=5pt,
  columns=flexible}
\lstdefinelanguage{yapscript}{
  morekeywords={
    Settings, outputDepth, endangermentDepth, mitigationDepth,
    simulationLength, suppressMishaps, suppressEndangerments,
    suppressMitigations, suppressResumptions, suppressStateLabel,
    suppressMishapLabel, suppressEndangermentsLabel,
    suppressMitigationsLabel, requiredVersion, extFile,
    graphDirection, grayscale, riskDiscount, allFactorsDirect,
    for, alias, desc, description,
    Activity, Task, enable, include, successor, initialState,
    ControlLoop, Controller, Plant, Application, Process, item, asset, device, partOf,
    poweredBy, embodiedBy, implementedBy,
    mode,event,sync, function,capability, update,safetyfun,target,filter, guard,
    pre, inv, post, constraint, state, type, role, class, hasFunction, 
    FactorModel, HazardModel, offRepair, direct, activatedBy, detectedBy,
    mitigatedBy, resumedBy, alleviatedBy, requires,
    requiresOneOf, requiresNOf, requiresMit, requiresOcc, requiresNot,
    causes, permits, excludes, prevents,
    preventsMit, mitPreventsMit, final, mishap, accident, incident,
    triggeredBy, impacts, phaseinv,
    distances, weights,
    AGENT, CONTROLLER
  },
  morecomment=[l]{//},
  morecomment=[l]{\#},
  morecomment=[s]{/*}{*/},
  sensitive=false}
\lstdefinestyle{yapscript}{
  frame=single,
  mathescape=true,
  breaklines=true,
  basicstyle=\srcfont,
  keywordstyle=\ttfamily\color{blue}\bfseries,
  identifierstyle=\ttfamily\slshape,
  stringstyle=\ttfamily,
  commentstyle=\rmfamily\color{gray},
  emphstyle=\ttfamily\color{red},
  tabsize=4,
  numbers=left,
  numberstyle=\tiny,
  stepnumber=2,
  numbersep=5pt,
  columns=flexible}
\newcommand{\mitem}[1][1]{\tikz[baseline=(#1.base),every
  node/.style={transform shape}, scale=.7] \node[stlab] (#1) {#1};}

\usepackage{wrapfig}
\usepackage{underscore}
\newcommand{\srcfont}{\footnotesize\ttfamily}

\providecommand{\event}{FMAS 2020} %
\usepackage{breakurl}             %
\usepackage{underscore}           %

\title{\Yap: Tool Support for Deriving Safety Controllers\\
  from Hazard Analysis and Risk Assessments}
\author{Mario Gleirscher\thanks{This research was funded by the Lloyd's
      Register Foundation under the AAIP grant CSI:Cobot.}
  \institute{Dept. of Computer Science,
    University of York,
    York, U.K.}
  \email{mario.gleirscher@york.ac.uk}}
\def\titlerunning{Safety Controllers with \Yap}
\def\authorrunning{M. Gleirscher}

\begin{document}
\maketitle

\begin{abstract}
  Safety controllers are system or software components responsible for
  handling risk in many machine applications.  This tool paper
  describes a use case and a workflow for \Yap, a research tool for
  risk modelling and discrete-event safety controller design.  The
  goal of this use case is to derive a safety controller from hazard
  analysis and risk assessment, to define a design space for this
  controller, and to select a verified optimal controller instance
  from this design space.  We represent this design space as a
  stochastic model and use \Yap for risk modelling and generation of
  parts of this stochastic model.  For the controller verification and
  selection step, we use a stochastic model checker.  The approach is
  illustrated by an example of a collaborative robot operated in a
  manufacturing work cell.
\end{abstract}

\section{Introduction}
\label{sec:introduction}

To ensure their safe operation, machines, such as mobile robots or
delivery drones, incorporate controllers responsible for the
\emph{handling of \acp{CE}} while performing their tasks.  We will
refer to such controllers as \emph{safety controllers}.  \Acp{CE} can
be failures, human errors, or other hazards, any causes thereof and
any consequences, such as incidents or accidents.  \ac{CE} handling
can involve the anticipation and mitigation of hazards and the
prevention and alleviation of accidents, \eg by switching a machine
into a mode with lower risk, \ie a \emph{safety mode}, by performing a
\emph{safety function}~(\egs a warning signal), or by changing the
machine's activity.  Therefore, safety controllers have to be
carefully specified, designed, and verified in order to be deployed
according to state-of-the-art regulations~\cite{ISO15066,IEC61508}.

This tool paper supplements our approach to the verified synthesis of
safety controllers~\cite{Gleirscher2020-SafetyControllerSynthesis}
with a hands-on guide to the research tool \emph{\Yap Against
  Perils}~\cite{Gleirscher-YapManual}.
One objective of \Yap is to support the steps required to transform
results from hazard analysis into verifiable models of safety
controllers.  \Yap seeks to bridge the gap between the identification
of hazards, the formulation of safety goals, and the implementation of
safety controllers.  Although \Yap might be more widely used, adaptive
and autonomous cyber-physical systems with their highly automated and
complex safety
mechanisms~\cite{Calinescu2018-EngineeringTrustworthySelf} are in its
focus.

\Cref{sec:example} briefly revisits the example discussed in more
detail in \cite{Gleirscher2020-SafetyControllerSynthesis}.
\Cref{sec:modelling-with-yap} describes preliminaries of \Yap.
\Cref{sec:overview} proposes a workflow to derive a safety controller.
\Crefrange{sec:step1-process-model}{sec:verif-contr-synth} detail
this workflow in the format of a hands-on guide.
\Cref{sec:outlook,sec:conclusions} discuss directions for future work
and conclude.

\section{Running Example: A Collaborative Manufacturing Robot}
\label{sec:example}

We illustrate the proposed workflow by example of a \ac{HRC} in a
manufacturing work cell with a collaborative robot
arm~\cite{Gleirscher2020-SafetyControllerSynthesis}.  This work cell
has a safeguarded workbench, which is manually supplied with work
pieces to be processed by a robot arm and a welder within a
safeguarded area next to the workbench.  The robot moves to the
workbench, grabs the work piece, and moves to the welder.  The robot
and the welder together perform a specific welding task on the work
piece.  After finishing this task, the robot arm returns the work
piece to the workbench where the operator picks it up and supplies the
robot with another work piece to repeat this cycle.  The work cell is
equipped with several safety modes~(\egs safety-rated monitored stop)
and safety functions~(\egs a warning display) operated by a safety
controller on occurrence of a \ac{CE}~(\egs operator close to weld
spot while welder and robot are working).  This way, the safety
controller works on top of this cyclic manufacturing process.

\section{Overview of \Yap: Modelling Concepts and Tool Features}
\label{sec:modelling-with-yap}

\Yap is a research tool for risk modelling, analysis, and 
design of safety controllers.\footnote{\Yap, its manual, and the
  running example can be obtained from
  \url{http://www.gleirscher.de/yap/} or from the author.}  \Yap's
input language is a \ac{DSL} providing a corresponding set of
modelling primitives.

\emph{Activities}
provide a finite abstraction of the physical process of interest and
can be useful for modelling the task structure of an application as
well as for structuring a risk analysis accordingly.  \Eg the task of
the robot arm exchanging a work piece can be separated from a welding
task performed by the welder and the robot arm.

\emph{Risk factors}~(factors for short) describe the \emph{life-cycle}
and constituents of \acp{CE} potentially being observed when
performing the activities, \eg the robot arm and the operator being on
the workbench simultaneously.  The hazard list can be modelled as a
list of factors.  \emph{Factor dependencies} specify temporal or
causal relations between risk factors~(\egs\yapkeyw{requires},
\yapkeyw{prevents}).  \Eg the fact that the robot arm touches the
operator \yapkeyw{requires} the operator to be in one of the
safeguarded areas.

A factor is modelled by four life-cycle \emph{phases}~(\ies
inactive $\inact$, active $\activ$, mitigated $\mitig$, and mishap
$\mishp$) and five \emph{events}~(\ies endangerment,
mitigation, resumption, mishap, alleviation).  Four of these events
can be refined into \emph{modes} and attributed with 
(quantitative) \emph{parameters}.  \Eg the \emph{endangerment} from an
operator entering the workbench while the robot is handling a work
piece there is \emph{detected} by a light barrier and the robot
position.  This event can be \emph{mitigated} by a safety-rated
monitored stop~(\texttt{srmst}) and signalling the operator to leave
the workbench.  After the operator has followed this advice, the robot
can \emph{resume} its work piece handling.  In case of a detected
mishap, a potential consequence could be \emph{alleviated} by a
complete shutdown of the work cell and an emergency call.  Modes can
be embodied by physical and logical \emph{items}, particularly,
\emph{actors} (synonymously, agents), constituting the
\emph{application}.  \Eg the logic for the safety-rated monitored stop
could be embodied by the robot arm.

These modelling primitives will be explained and used in
\Cref{sec:guide-contr-design}.  A more detailed description of \Yap's
\ac{DSL} is, however, provided in \cite{Gleirscher-YapManual}.
Overall, \Yap models can inform controller design by injection of a
model of a safety controller into a process model of the application.
However, apart from this use case, with \Yap's \ac{DSL} one can
describe operational risk as an abstract state machine, explore its
symbolic state space, \ie the \emph{risk space}, shape its transition
relation, and perform a light-weight symbolic simulation of \ac{CE}
occurrence and handling.  Furthermore, one can calculate risk spaces
and properties thereof~(\egs mitigation orders
\cite{Gleirscher2020-RiskStructuresDesign}), and generate \aclp{MCSeq}
\cite{Gleirscher2020-SafetyControllerSynthesis,Gleirscher-YapManual}.

\section{Overview of the Workflow}
\label{sec:overview}

\begin{figure}
  \centering
  \includegraphics[width=\textwidth]{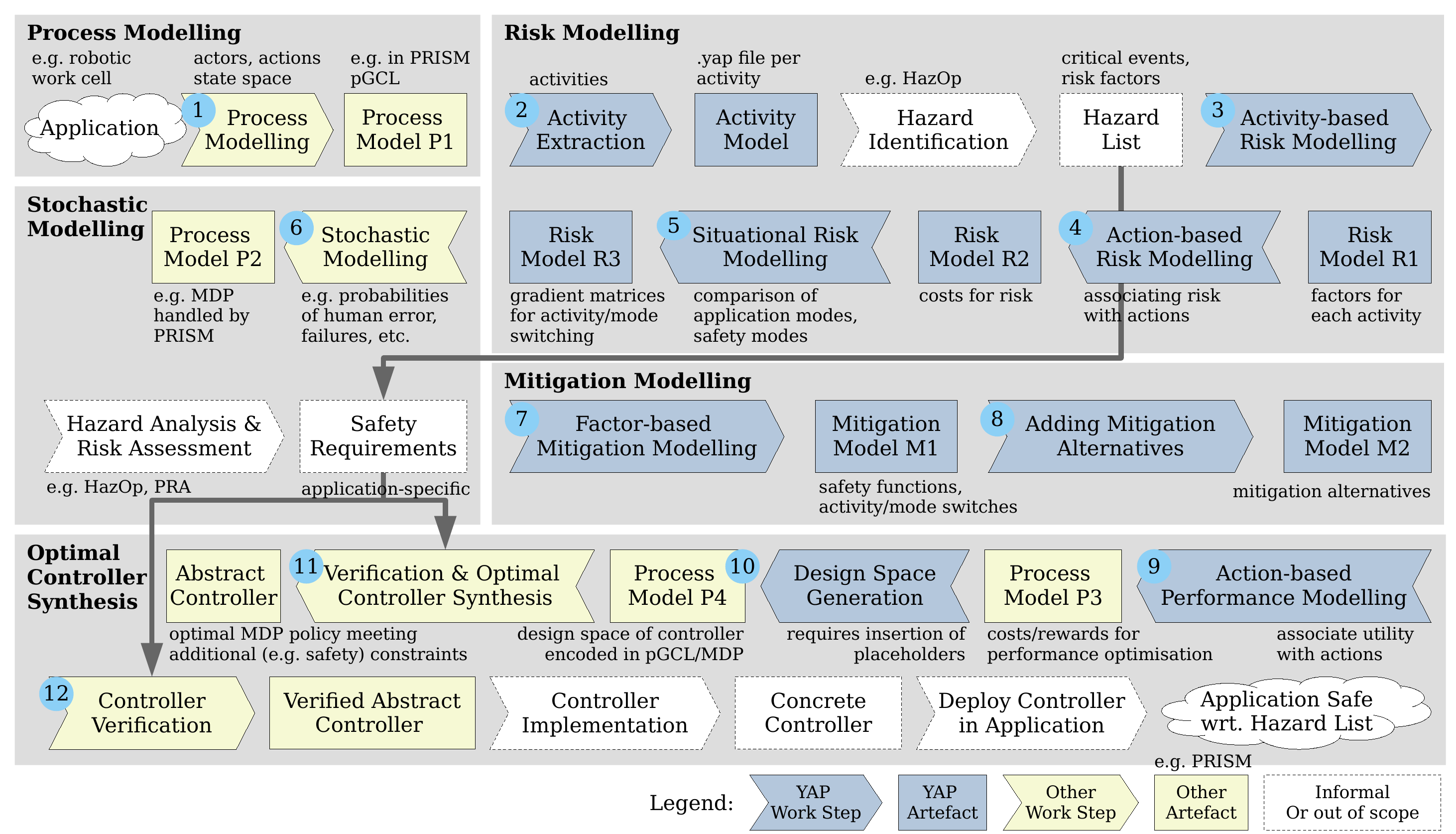}
  \caption{Intended workflow to be used with \Yap and the stochastic
    model checker
    \label{fig:yap-workflow}}
\end{figure}

\Cref{fig:yap-workflow} indicates the methodological context in which
\Yap can be used.  There, the specification and synthesis of a safety
controller consists of several work steps (\mitem[1] to \mitem[12]) in
five stages.

\paragraph{Process Modelling.}

\mitem[1] One begins with constructing a behavioural model (P1) of the
physical process of interest, focusing on actors, actions, and the
state space these actions can modify~(\Cref{sec:step1-process-model}).
For this use case of \Yap, we encode this model in the \ac{pGCL} of
the \Prism stochastic model
checker~\cite{Kwiatkowska2011-PRISM4Verification} for the analysis of
\acp{MDP}.  An \ac{MDP} is a stochastic model, particularly
well-suited for reasoning about processes with non-deterministic
decisions over actions and probabilistic outcomes of these actions.
\ac{pGCL} offers a concise way of encoding \acp{MDP}.  \Prism
conveniently implements \ac{pGCL} with %
parallel composition~\cite{Hoare1985-CommunicatingSequentialProcesses,
  Parker2019-PRISMModelChecker}.

\paragraph{Risk Modelling.}

\mitem[2] To structure the risk analysis, we
decompose the process into activities~(\Cref{sec:primer-yap}).
\mitem[3] For each activity, we identify risk factors and incidents
using an appropriate technique~(\egs the hazard identification stage
of \ac{HazOp} \cite{IEC61882}) and capture the results, \ie the hazard
list, in a risk model (R1, \Cref{sec:step2-activity-factor-models}).
The practical reasoning and the domain expertise to apply when using a
typical system \ac{HARA} technique, are very well-documented in a rich
corpus of literature~(\egs \cite{Leveson2012-EngineeringSaferWorld,
  Knight2012-FundamentalsDependableComputing,
  Ericson2015-HazardAnalysisTechniques}).  Hence, we touch \ac{HARA}
specifics only selectively.
\mitem[4] We extend this risk model (R2) by associating to every
action a number that encodes the risk of any of the captured incidents
from the performance of this particular
action~(\Cref{sec:step6-optimal-process}).
\mitem[5] Furthermore, the risk model (R3) is extended by assigning
to every transition between two activities or safety modes a number
that encodes the increase or decrease in risk of any incident when
taking this transition~(\Cref{sec:step4-gradient-matr}).  The result
are two \emph{gradient matrices}, one for activity changes and one for
safety mode changes.

\paragraph{Stochastic Modelling.}

\mitem[6] We return to the process model and introduce probabilistic
phenomena such as human error and sensor
failure~(\Cref{sec:integr-prob-into}).  As a result, we get a process
model (P2) amenable to \ac{HARA}, \eg further steps of \ac{HazOp} and, particularly,
\ac{PRA}.

\paragraph{Mitigation Modelling.}

\mitem[7] Based on the process and risk models, we design mitigations
for each of the identified risk factors~(\Cref{sec:step3-mode-specs}).
An individual mitigation can perform a safety function, an activity
change, and safety mode change, and will after removal of the
corresponding risk factor return the process to a state where normal
operation can continue.
Within \Yap, the mitigation model (M1) refines the risk model by an
abstract state machine.   This state machine is translated into the
language used for the process model.
\mitem[8] \Yap allows the definition of alternatives for the
mitigation of a single risk factor~(\Cref{sec:step5-optimal-mitig}).
The result is a mitigation model (M2) with these alternatives creating
a \emph{controller design space}.

\paragraph{Verified Controller Synthesis.}

\mitem[9] Similar to the assignment of risk to process actions,
we now associate other costs and rewards~(\egs nuisance of the
operator, energy consumption, utility of a performed robot action)
with these actions~(\Cref{sec:acti-based-perf}).  From this step, we
obtain a reward-enhanced \ac{MDP} model of the application (P3).
\mitem[10] Next, we use \Yap's \ac{pGCL} generator to translate the
risk and mitigation models into a set of \ac{pGCL} fragments.  These
fragments fill placeholders easily inserted into the process model
beforehand.  The resulting process model (P4) is amenable to property
verification and acts as a design space for controller
synthesis~(\Cref{sec:step7-synth-mdp}).  This design space includes
the set of choice resolutions of the \ac{MDP} as well as further
degrees of freedom stemming from other unfixed model parameters~(\egs
probabilities \mitem[6], rewards \mitem[9]).  We can now verify
properties of this design space.
\mitem[11] Importantly, we use \Prism not only for property
verification but also for the selection of a policy~(also called
adversary or strategy), \ie a particular choice resolution
representing the controller, from this design space.  Optimal \ac{MDP}
policies are artefacts of quantitative verification and can be
represented as \acp{DTMC}.  Depending on the 
safety requirements, the chosen policy will have to meet certain
safety constraints and be Pareto-optimal \wrt the considered
performance criteria~(\Cref{sec:step8-synth-dtmc}).
\mitem[12] We finally verify further safety properties of the
policy~(\Cref{sec:policy-verif}).
Theoretically, \mitem[11] and \mitem[12] could be collapsed into one
verification step carried out on the design space.  However,
the tooling for this use case requires us to separate
Pareto optimisation and constraint verification.

\section{Workflow for Controller Design}
\label{sec:guide-contr-design}

The following sub-sections provide a hands-on guide detailing the
workflow outlined in \Cref{fig:yap-workflow}.

\subsection{\mitem[1] Process Modelling: The Physical World}
\label{sec:step1-process-model}

We create a stochastic model~(an \ac{MDP}) of the
manufacturing process, as described in \Cref{sec:example}, and employ
the \Prism model checker for its analysis.  We model
\emph{actors}~(\egs a robot arm, an operator), \emph{activities}~(\egs
\yapact{exchWrkp} for exchanging a work piece, \yapact{welding} a work piece), and~(atomic)
\emph{actions}~(\egs the robot grabs the work piece, the operator
enters the work cell).  In \Prism, actors can be implemented as
modules and actions as guarded commands of the form \texttt{[EVENT]
  GUARD $\rightarrow$ UPDATE}.

\Cref{lst:processmodel} exemplifies the actor \yapag{robotArm},
participating in the two activities \yapact{exchWrkp} and
\yapact{welding} with several actions.  \Eg the involvement of
\yapag{robotArm} in \yapact{exchWrkp} is implemented by the three
actions \yapcmd{r\_moveToTable}, \yapcmd{r\_grabLeftWorkpiece}, and
\yapcmd{r\_placeWorkpieceRight}.
The structure of many of the modelled actions follows a specific
pattern:
\[
  \mathtt{[ACTOR\_ACTION]\; !CYCLEEND\; \&\; SM\; \&\; ACTIVITY\; \&\;
  CUSTOM\; \rightarrow\; UPDATE} \;.
\]
Action names carry prefixes to indicate the actor(s) performing these
actions, \ie \textbf{r} for a robot action, \textbf{rw} for a compound
action of the robot and the welder, \textbf{h} for a physical human
action, \textbf{hi} for an internal human decision, \textbf{s} for a
synchronous action of the safety controller and other actors,
\textbf{si} for an independent controller action.
\texttt{CYCLEEND} is a predicate that, when \texttt{true}, terminates
model execution.  Each action has to be guarded by the
activities~(\texttt{ACTIVITY}) and safety modes (\texttt{SM}) it is allowed
to be performed in.  Action-specific guards~(\texttt{CUSTOM}) and
updates~(\texttt{UPDATE}) are conjoined and specified.  The
\yapag{robotArm} does not contain stochastic phenomena whereas
\yapag{humanOp} and other parts of the process model do.  The use of
probabilistic updates for human errors and other phenomena will be
further discussed in \Cref{sec:integr-prob-into}.

\begin{lstlisting}[float=t,language={prism},frame=none,
caption={Process model fragment for the robot arm in \Prism},
label=lst:processmodel, stepnumber=1]
module robotArm 
reffocc: 	bool 		init false; // is the grabber occupied?
wpfin: 		bool 		init false; // is the work piece finished?
rloc: 		[atTable..atWeldSpot] 	init inCell; // robot arm location
//<%
...
// exchWrkp: exchange a work piece between workbench and welder
[r_moveToTable] !CYCLEEND & (safmod=normal|safmod=ssmon|safmod=pflim) & ract=exchWrkp & !rloc=sharedTbl & ((wps!=right&reffocc)|wps=left&!reffocc) -> (rloc'=sharedTbl);
[r_grabLeftWorkpiece] ...;
[r_placeWorkpieceRight] ...; ...
// welding: carry out welding task together with welder
[r_moveToWelder] !CYCLEEND & (safmod=normal|safmod=ssmon|safmod=pflim) & ract=exchWrkp & reffocc & !wpfin -> (ract'=welding)&(rloc'=atWeldSpot);
[rw_weldStep] ...;
[rw_leaveWelder] ...; ...
endmodule
\end{lstlisting}

\subsection{Risk Modelling with \Yap}
\label{sec:contr-deriv-with}

\paragraph{\mitem[2] Activity Modelling.}
\label{sec:primer-yap}

\begin{wrapfigure}[5]{r}{4cm}
\begin{lstlisting}[aboveskip=-1.2em,]
Activity {
  include moving;
  successor welding;  
  successor off;      
  successor idle;     
}
\end{lstlisting}
\end{wrapfigure}

\begin{figure}
  \centering
  \includegraphics[width=.9\textwidth]{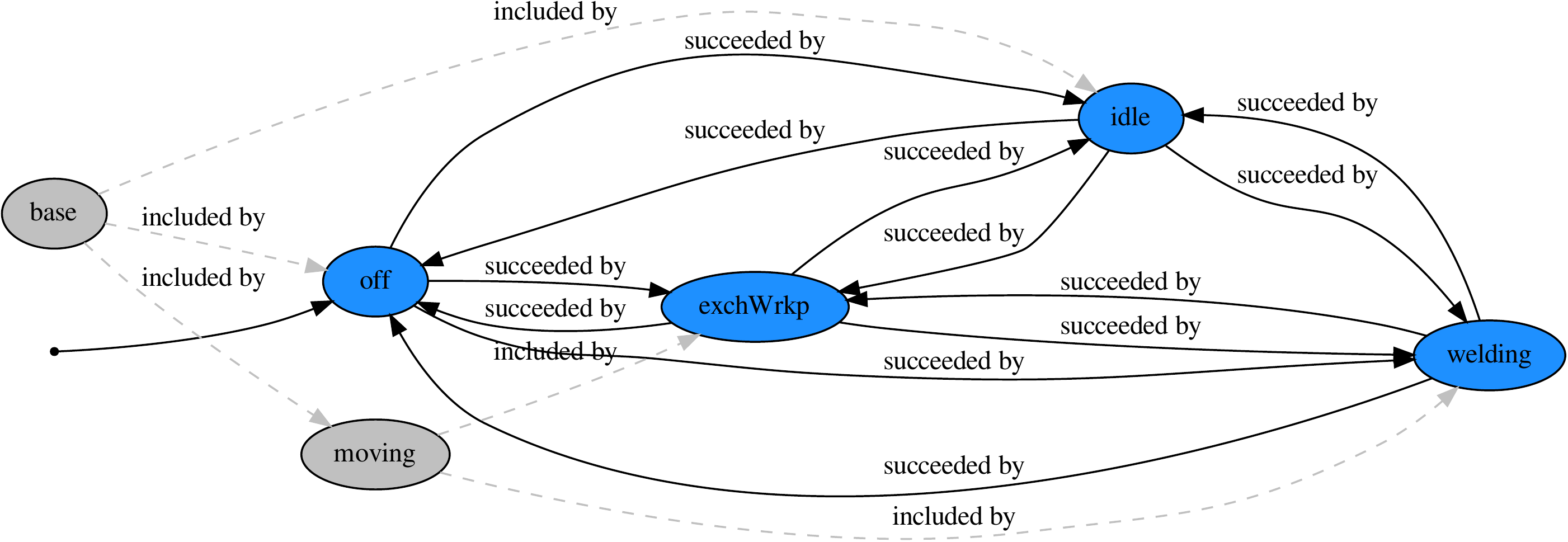}
  \caption{Activity graph showing all activities reachable from
    the activity \yapact{off} 
    \label{fig:exa-hrc-off}}
\end{figure}

We create a \Yap file for each activity~(\egs generic task or
sub-task) of the manufacturing process.  \Eg for the activity
\yapact{exchWrkp}, the listing on the right specifies relationships to
other activities.
Particularly, \yapact{exchWrkp} inherits~(\yapkeyw{includes})
attributes~(\ies hazard, activity successors, \etc) from
the generic activity \yapact{moving} and can be followed~(\yapkeyw{successor}) by either
of the basic activities \yapact{off}, \yapact{welding}, or \yapact{idle}.
With the following command, \Yap identifies all activities reachable from
the activity \yapact{off} in form of a \ac{LTS}.  For inspection, this
\ac{LTS} is visualised as a graph
in \Cref{fig:exa-hrc-off}.
\begin{verbatim}
  yapp --global-logging -m off.yap \
       -o output/off-act.dot --showmodel activities .
\end{verbatim}

\paragraph{\mitem[3] Activity-based Risk Modelling.}
\label{sec:step2-activity-factor-models}

We create a hazard list for each activity, \eg by performing a
\ac{HazOp}~\cite{IEC61882}.  Then, we derive factor specifications
from these lists.  For instance, for the activity \yapact{welding}, we
specify five factors, such as the factor ``\textsf{H}uman arm and
\textsf{R}obot on shared \textsf{W}orkbench''~($\rfct[HRW]$,
\Cref{lst:factor-HRW-RT}).  Then, we specify factor dependencies, \eg
$\rfct[HRW]$ \yapkeyw{requires} the factor ``\textsf{H}uman arm on
\textsf{W}orkbench''~($\rfct[HW]$), \yapkeyw{prevents} the activation
of factor ``\textsf{H}uman \textsf{C}lose to weldspot''~(\textsf{HC}),
and $\rfct[HRW]$'s mitigation prevents the
mitigation~(\yapkeyw{mitPreventsMit}) of $\rfct[HC]$ and
``\textsf{H}uman in \textsf{S}afeguarded area''~($\rfct[HS]$).

The inclined reader will recognise that $\rfct[HRW]$ is not only a
critical event in the process, it also contains the hazard
``\textsf{R}obot on shared \textsf{W}orkbench''.  As a whole,
$\rfct[HRW]$ can be seen as a latent cause of the incident
``\textsf{R}obot arm \textsf{T}ouches the operator''~($\rfct[RT]$).
Accordingly, the specification for $\rfct[RT]$ in
\Cref{lst:factor-HRW-RT} uses the constraint
\lstinline[emph={HRW}]+requiresNOf (1|HRW,HS,HC)+
to refer to its potential causes: at least one of $\rfct[HRW],
\rfct[HS]$, and $\rfct[HC]$.

\begin{figure}
\begin{lstlisting}[emph={HRW,RT},multicols=2]
  HRW desc "(H)uman arm and (R)obot on 
            shared (W)orkbench"
    requires (HW)
    prevents (HC)
    mitPreventsMit (HS,HC) ...
  RT desc "(R)obot arm (T)ouches 
          the operator"
    requiresNOf (1|HRW,HS,HC)
    mitPreventsMit (RC) 
  ...
\end{lstlisting}
\caption{Specifications of the two risk factors {$\rfct[HRW]$} and
  $\rfct[RT]$ \label{lst:factor-HRW-RT}}
\end{figure}

\paragraph{\mitem[4] Action-based Risk Modelling.}
\label{sec:step6-optimal-process}

In \Yap, actions of both the controlled process and the safety
controller can be characterised using multiple weights that order the
choice alternatives when numerical solvers search the \ac{MDP} policy
space.  \Yap converts these weights into action reward structures used
for quantitative verification in tools such as \Prism.

Like for the actions of the safety controller, one can specify such
characteristics for the actions of any actor in the process.  This is
done in \Yap by providing a \yapkeyw{weight}
structure~(\Cref{yap:weights-risk}).  For action-based risk modelling,
parameters prefixed with \texttt{risk\_}, say \texttt{risk\_}$\rfct$
for a factor $\rfct$, are used to accumulate risk in reward
calculations for the underlying \ac{MDP}.  The corresponding action
rewards are guarded by $\activ$, requiring that $\rfct$ is active.
\Eg if \yapkeyw{weights} provides a \texttt{risk\_}$\rfct$ entry for
the process action \yapcmd{a} then \yapcmd{a} is rewarded only if it
is taken in a state where $\activ$
holds~\cite{Kwiatkowska2011-PRISM4Verification}.

\begin{lstlisting}[basicstyle=\footnotesize\ttfamily,float,multicols=2,
caption={Structure for specifying risk values on a per-action basis},
label=yap:weights-risk]
Weights rewards {     guard      risk_HC;
  // robotArm
 r_moveToTable:       ""         "5";
 ...
  // welder
 rw_weldStep:         ""         "10"; 
 rw_leaveWelder:      ""         "5";
  // humanOp
 ...
 h_approachWeldSpot:  ""         "7";  
 h_exitCell:    "notif=leaveArea" "0"; }
\end{lstlisting}

\paragraph{Connecting the \Yap and \Prism Models.}

Crafting the \Yap model and generating a controller model
to be integrated into a process model require the engineer
to have substantial knowledge of the process model.
\Eg in order for \Yap to inject valid synchronisation commands into
existing \Prism modules, one has to specify the activities the
respective actors are involved in.\footnote{That could also be
  achieved by parsing the process model~(here, the \Prism file) but is
  beyond \Yap's current functionality.}  This is done by providing a
\texttt{validActs} parameter taking values of the form
\texttt{act1|act2|...|actn}.  \Eg in \Cref{lst:model-conn}, the
\yapag{robotArm} is involved in the three activities
\yapact{exchWrkp}, \yapact{welding}, and \yapact{off}, the
\yapag{welder} in \yapact{welding}, \yapact{idle}, and \yapact{off}.
If \texttt{validActs} is not provided then \Yap assumes that the actor
can be involved in any of the specified activities.
Furthermore, additional module-specific~(\ies local) variables used by
the safety controller can also be declared as part of an item
specification.

\begin{lstlisting}[float,emph={validActs,notif,notif_leaveWrkb},multicols=2,
caption={Connecting \Yap and \Prism models via \Yap agents and \Prism modules},
label=lst:model-conn]
  robotArm type AGENT
    validActs="exchWrkp|welding|off";
  welder type AGENT
    validActs="welding|idle|off";
  safetyCtr type CONTROLLER
    notif="[ok..resetCtr] init ok" 
    notif_leaveWrkb="bool init false";
\end{lstlisting}

\paragraph{\mitem[5] Situational Risk Modelling.}
\label{sec:step4-gradient-matr}

In \Yap, we can not only model risk in terms of factors but also in
terms of the behavioural modes the application or controlled process
can be in.  Examples of such modes are activities and safety
modes~(\Cref{sec:step3-mode-specs}) as used and standardised in
application domains such as \ac{HRC}~\cite{ISO15066}.  Of course, we
allow behavioural modes to change during operation.  Moreover, we
often do not exactly know about the absolute risk level of a certain
mode in a certain situation.  Thus, \Yap offers the possibility to
define \emph{risk gradient matrices}\index{risk gradient matrix} that
only capture expected changes in the risk level when changing from one
mode into another.  We only consider symmetric changes and, thus, use
skew-diagonal matrices reduced\footnote{For readability, one
  can provide the symmetric upper right triangle as well.  However,
  \Yap internally mirrors the values from the lower left to the upper
  right to ensure skew-diagonality.}  to their lower left
triangles.  Examples of these matrices are shown in
\Cref{lst:grad-matr}.

\begin{lstlisting}[float,multicols=2, frame=none, emph={act,safmod},
caption={Two risk gradient matrices},
label=lst:grad-matr]
Distances act {
 off:       0;
 idle:      1 0; 
 exchWrkp:  3 2 0;
 welding:   5 4 2 0; }      
Distances safmod {
 normal:   0;
 hguid:   -2  0;
 ssmon:   -1  1    0;
 pflim:   -2  0   -1  0;
 srmst:   -3 -1   -2 -1  0;
 stopped: -4 -2   -3 -2 -1  0; }
\end{lstlisting}

To use such matrices for maximisation in \Yap's \ac{pGCL}
generator,\footnote{A description of \Yap's algorithms is out of scope
  of this tool paper.}  we associate a positive gradient with an
improvement of the risk level and, vice versa, a negative gradient
with a worsening of the risk level.\footnote{This choice should not be
  too counter-intuitive as one can associate negative numbers with
  something undesirable.}  \Eg the \yapcmd{act} matrix tells us that a
transition from the activity \yapact{welding} to the activity
\yapact{exchWrkp} improves the risk level by $2$.  As a further
example, the \yapcmd{safmod} matrix states that a transition from the
safety mode \yapact{srmst}~(\ies safety-rated monitored stop) to the
safety mode \yapact{ssmon}~(\ies speed and separation monitoring)
worsens the risk level by $-2$.

\Yap uses these matrices to calculate \yapcmd{act}- and
\yapcmd{safmod}-updates of the set of guarded commands generated for
the safety controller from the given factor
specifications~\cite{Gleirscher2020-SafetyControllerSynthesis}.
In \Yap, risk can be evaluated both based on transitions
through the risk space~(\ies from one risk state to another) and on a
wider situational basis~(\ies based on transitions from one activity
or safety mode to another).  Note that while in step \mitem[4],
actions are associated with an \emph{absolute risk} value, in step
\mitem[5], mode and activity changes are associated with a \emph{risk
  gradient}, a relative measure stemming from a pair-wise comparison
of modes and activities.  Whereas for individual actions, the \Yap user
needs to agree on a single global scale for risk assessment, the higher
complexity of risk assessment for modes and activities is taken account
of by gradients.

\subsection{\mitem[6] Stochastic Modelling}
\label{sec:integr-prob-into}

Probabilistic choice in the process model set up in
\Cref{sec:step1-process-model}~(\egs an \ac{MDP} modelled in \Prism's
\ac{pGCL}) can capture a variety of adverse and critical stochastic
events, such as human errors, sensor failures, actuator perturbation
and failures, and mishaps.

\paragraph{Human Errors.}

We consider, \eg intrusion of \yapag{humanOp}, whether
intentional or erroneous, into the work cell when not allowed:
\begin{lstlisting}[frame=none,language={prism},emph={hi_mayEnterCell},]
[hi_mayEnterCell] !dntFlg_enterCell & !mntDone & !CYCLEEND 			
	// deontic flag and end-of-cycle check
	& wps!=empty & hact=idle & (hloc!=inCell & hloc!=atWeldSpot) & !wpfin 	
	// action physically possible/feasible/reasonable
	-> ((prm_enterCell&req_enterCell)?.9:.2):(dntFlg_enterCell'=true)	
	// action enabled if(90)/ifnot(10) allowed/required
	  +((prm_enterCell&req_enterCell)?.1:.8):true;				
	// action notenabled if(10)/ifnot(90) allowed/required
\end{lstlisting}

This listing shows a two-staged\footnote{Such commands do not increase
  the expressiveness of \ac{pGCL} as they can be expressed by a set of
  ordinary commands.  However, they increase convenience by allowing
  what can be called conditional probabilistic choice.} guarded
command modelling the operator's internal decision, not their physical
action.  The flag \texttt{dntFlg\_enter\-Cell} is \emph{deontic} in
the sense that it \emph{enables but not triggers} the action
\yapcmd{h\_enterCell} of the operator actually entering the work cell.
The activation of this flag is handled by a \emph{conditional
  probabilistic choice}: the condition \texttt{prm\_enterCell}
distinguishes between probabilistic intrusion in states where the
human operator is \emph{permitted} to enter the work cell and
probabilistic intrusion where the operator is not allowed to enter.
In case of permission, \texttt{dntFlg\_enterCell} is set to
\texttt{true} with a 90\% chance and, in case of denial, only with a
20\% chance.
This is a way of saying that the operator commits a human error in
20\% of the times \emph{when they should not} enter the work cell.
The predicate \texttt{req\_enterCell} is not used in this particular
model.  It just indicates another potentially useful state selection
mechanism.  The condition and the deontic flag can be omitted if the
probabilities are universal and we do not need to separate a
\emph{physical} action~(\egs movement of the operator) from a
\emph{logical} action~(\egs change of the operator's mind).
In this model, we treat operating errors and malicious misuse in the
same way, however, a distinction can be necessary in other contexts.

\paragraph{Sensor Failures.}

The separation of physical and logical human actions allows us to
synchronise sensor actions with physical actions, \eg to model a
range detector with a 5\% chance of failure:
\begin{lstlisting}[frame=none,language={prism},emph={h_enterCell},numbers=none]
[h_enterCell] true -> .95:(rngDet'=near)+.05:true;      
\end{lstlisting}

\yapcmd{h\_enterCell} is an event with two
\emph{synchronous} actions, the 
physical action of \yapag{humanOp} entering the work cell and the
logical action of the range finder in \yapag{sensorUnit}.
Synchronisation establishes real-time behaviour despite the deontic
nature of \ac{pGCL}.  This sensor failure is modelled by the range
finder signalling \texttt{near} to \yapag{safetyCtr} only in 95\% of
the cases where \yapag{humanOp} actually enters the cell.

\paragraph{Actuator Perturbation and Failures.} %

We have not modelled any perturbations in the \ac{HRC} case study.
However, an obvious entry point for such phenomena would be the
actions of \yapag{robotArm}.  \Eg the action
\yapcmd{r\_grabLeftWorkpiece} could be extended by a probabilistic
update modelling the fact that grabbing a work piece fails in a
certain fraction of trials with a work piece not in the grabber and/or
still in the work piece support, or whatever outcome seems realistic.

\paragraph{Mishaps after Critical Events.}

Based on conditional probabilistic choice, similar to human error
modelling, we use probabilities to capture the fact that from
activated critical events and certain actions, a transition into a
mishap state is possible.  Consider the example:
\begin{lstlisting}
  HC desc "(H)uman (C)lose to active welder and robot working"
    ...
    mis="h_exitCell" // mishap possibly initiated by the action h_exitCell
    prob=0.05 // probability of mishap is 5 percent if unrecognised OR active and not mitigated
    sev=5; // severity of the mishap is of class 5
\end{lstlisting}
The parameter \texttt{prop=0.05} defines the probability of a
\texttt{mis}hap from the action \yapcmd{h\_exitCell} in case of an
activated $\rfct[HC]$ to be 5\%.  The parameter \texttt{sev=5}
associates the impact from such a mishap to the exemplary impact class
5, which can, \eg mean ``high''.  In other words, if \yapag{humanOp}
wants to perform the action \yapcmd{h\_exitCell} if the factor
$\rfct[HC]$ is active~(\ies $\activ[HC]$) then, with a 5\% chance, the
outcome will be the mishap $\mishp[HC]$, a specimen of the
\texttt{MISHAP} state.  The resulting \ac{pGCL} fragment generated by \Yap:
\begin{lstlisting}[frame=none,language={prism},
emph={CE_HC,CE_HRW,h_exitCell,h_placeWorkpieceLeft}]
[h_exitCell] true -> 
	((!HCp=mis & (CE_HC | RCE_HC))?0.05:0):(HCp'=mis)
	+((!HCp=mis & (CE_HC | RCE_HC))?0.95:1):true;
[h_placeWorkpieceLeft] true -> 
	((!HRWp=mis & (CE_HRW | RCE_HRW))?0.01:0):(HRWp'=mis)
	+((!HRWp=mis & (CE_HRW | RCE_HRW))?0.99:1):true;
\end{lstlisting}
\noindent
$\mishp[HC]$ can, for instance, be a welding spark injuring the
operator, or the \yapag{robotArm} hitting them, both encoded in a
corresponding risk/severity reward generated by \Yap:
\begin{lstlisting}[frame=none,language={prism},numbers=none,
emph={CE_HC,CE_HRW,h_exitCell,h_placeWorkpieceLeft}]
[h_exitCell] (!HCp=mis & (CE_HC | RCE_HC))	: 5.0;
\end{lstlisting}

One downside of this way of reward modelling in \Prism is that the
reward is paid independent of the outcome of \yapcmd{h\_exitCell}.  A
solution not chosen here would be to introduce an intermediate state
with the disadvantage of doubling the state space each time such a
construction is chosen.

\subsection{Mitigation Modelling with \Yap}
\label{sec:mitig-model}

\paragraph{\mitem[7] Factor-based Mitigation Modelling.}
\label{sec:step3-mode-specs}

The basic factor model allows mitigation in one~(direct) or two
stages~(indirect).  For controller synthesis, we focus on the
two-staged approach, which suggests the specification of modes for
detection, mitigation, and resumption.\footnote{We omit alleviation
  modes for the sake of simplicity of this guide.}  \Eg the modes for the factor
$\rfct[HRW]$ are referred to from its specification:

\begin{lstlisting}[emph={HRWdet,HRWmit,HRWres}]
  HRW desc "(H)uman arm and (R)obot on shared (W)orkbench" ...
    guard "hACT_WORKING & rloc=sharedTbl & hloc=sharedTbl"
    detectedBy (SHARE.HRWdet)
    mitigatedBy (PREVENT.HRWmit)
    resumedBy (.HRWres) ...
\end{lstlisting}
\noindent
and detailed in the \yapkeyw{Application} fragment:

\begin{lstlisting}[emph={HRWdet,HRWmit,HRWres}]
  mode HRWdet 
    guard "hACT_WORKING & rloc=sharedTbl & lgtBar=true"
    embodiedBy cellObSys;
  mode HRWmit desc "safety-rated monitored stop"
    update "(notif_leaveWrkb'=true)" // safety function on
    target (safmod=srmst) // safety mode on
    embodiedBy robotArm
    disruption=5 nuisance=1 effort=0.5;
  mode HRWres
    guard "hloc!=sharedTbl" // checking for hazard removal
    update "(notif_leaveWrkb'=false)" // safety function off
    target (safmod=normal); // safety mode off
\end{lstlisting}

\noindent
Detection, mitigation, and resumption modes make use of the process
model.  \Eg \yapcmd{HRWdet} uses the formula
\texttt{hACT\_WORKING}~(explained below) and the module variables
\texttt{rloc} and \texttt{hloc}~(the locations of the robot arm and
the operator) in an embedded \yapkeyw{guard} expression.
\yapkeyw{update} in \yapcmd{HRWmit} specifies the assignment of
\texttt{true} to the module variable \texttt{notif\_leaveWrkb} to
notify the operator to leave the workbench.  Being a generic form of
update, \yapkeyw{target} specifies mode switching preferences that,
when unsafe, will be overridden by \Yap using the gradient
matrices~(\Cref{lst:grad-matr}).

\Eg \yapcmd{HRWmit} switches to the mode ``safety-rated monitored
stop''~(\texttt{smrst}) and \yapcmd{HRWres} back to the
\texttt{normal} mode.  However, \yapag{safetyCtr} would only switch to
the \texttt{normal} mode if this is acceptable in the risk state to be
reached.  Hence, each of these modes will be enhanced by information
about activities and risk states and translated into one or more guarded
commands for being integrated into the process model.  Note that there
are also \emph{embodiment} references to the items \yapitem{robotArm}
and \yapitem{cellObSys}, the latter being the overall sensor system of
the work cell.

\paragraph{Connecting the \Yap and \Prism Models.}

It is convenient to reuse formulas in several places.  For this
purpose, \Yap allows their definition in the \yapkeyw{Application}
sections of the \Yap model:
\begin{lstlisting}[emph={hACT\_WORKING,hFINAL\_CUSTOM,hST\_HOinSGA}]
Application cobot {
  hACT_WORKING = "(ract=exchWrkp | ract=welding | wact=welding) & safmod=normal"; 
  ...
  hST_HOinSGA = "hloc=inCell | hloc=atWeldSpot";
  hFINAL_CUSTOM = "wpfin & wps=empty & !reffocc & mntDone"; ... }
\end{lstlisting}

\noindent
\Eg the predicate \texttt{hACT\_WORK\-ING} includes activities where
actors are effectively working.  \texttt{hST\_HOinSGA} is a shortcut
specifying states where the operator is in the safeguarded area.
\texttt{hFINAL\_CUSTOM} refines
\texttt{CYCLEEND}~(\Cref{sec:step1-process-model}), the termination of
a process cycle, in our example, the end of a manufacturing cycle in
the work cell.  This predicate is used in the reduction of cyclic end
components in order for optimal \ac{MDP} policy search algorithms to work
correctly~\cite{Kwiatkowska2011-PRISM4Verification}.

\paragraph{\mitem[8] Adding Mitigation Alternatives.}
\label{sec:step5-optimal-mitig}

Factor specifications allow one to provide several mitigation and
resumption options and characterise their properties using \emph{risk
  and performance} parameters.  \Eg the factor
$\rfct[HS]$~(\textsf{H}uman in \textsf{S}afeguarded area while robot
working or welding) refers to three mitigation options in its
\yapkeyw{mitigatedBy} directive:
\begin{lstlisting}[emph={ssmon,srmst,stopped}]
  HS desc "(H)uman in (S)afeguarded area while robot working or welding"
    guard "hACT_WORKING & (hloc=inCell | hloc=atWeldSpot)"
    detectedBy (SHARE.HSdet)
    mitigatedBy (.ssmon,.srmst,.stopped)
    resumedBy (.HSres)
\end{lstlisting}

\noindent
We model the options \yapcmd{ssmon}, \yapcmd{srmst}, and
\yapcmd{stopped} in more detail in \Cref{lst:mit-opt}.
\begin{lstlisting}[float,emph={ssmon,srmst,stopped},
caption={All modes including for the factor
{$\rfct[HS]$} three mitigation options},
label=lst:mit-opt,multicols=2]
  mode HSdet desc "range detector"
    guard "hACT_WORKING & (rngDet=near | rngDet=close)";
  mode ssmon
    desc "speed/separation monitoring"
    target (safmod=ssmon)
    disruption=9 nuisance=9 effort=8;
  mode srmst
    desc "safety/rated monitored stop"
    target (safmod=srmst)
    disruption=5 nuisance=9 effort=5;
  mode stopped
    desc "protective emergency stop"
    target (safmod=stopped)
    disruption=2 nuisance=6 effort=3;
  mode HSres
    guard "!hST_HOinSGA" // check hazard removal
    target (safmod=normal);
\end{lstlisting}

\subsection{Verified Controller Synthesis}
\label{sec:verif-contr-synth}

\paragraph{\mitem[9] Action-based Performance Modelling.}
\label{sec:acti-based-perf}
Along with the three mitigation options specified in
\Cref{lst:mit-opt}, we provide estimates for their \emph{disruption}
of the manufacturing process, for their \emph{nuisance} of operators,
and for the \emph{effort} required for their execution.
In addition, for each action in the process
model~(\Cref{sec:step1-process-model}), one can provide a \emph{guard}
and several columns of optimisation parameters~(\egs
\texttt{prod}, \texttt{eff\_process\_time}, \texttt{risk\_HC}).  Each
such column is converted into an action reward structure.
\begin{lstlisting}[basicstyle=\footnotesize\ttfamily,multicols=2]
Weights rewards {
            guard    prod  eff_process_time; 
  // robotArm
 r_moveToWelder:  ""         "h"    "2*macro";
 ...
  // welder
 rw_weldStep: 	  ""         "h"    "3*macro"; 
 rw_leaveWelder:  ""         "h"    "macro";
  // humanOp
 h_start: 	  ""         "l"    "macro";
 ...
 h_enterCell:	  ""         "none" "none";
 h_exitCell: "notif=leaveArea""none""none"; 
}
\end{lstlisting}

\noindent
Moreover, as shown in these two examples, one can also use parameters
defined elsewhere~(\egs \texttt{macro, h, none}) instead of using
literal numbers in place.  One may provide several \yapkeyw{weights}
structures across the activity model.  However, they will all be
merged into one central ``database'' containing all columns found in
the given structures.

\paragraph{\mitem[10] Design Space Generation.}
\label{sec:step7-synth-mdp}

The process model has to be instrumented with \emph{placeholders} to
be substituted with model fragments generated by \Yap.  Such
placeholders take the form \yaptplparam{\emph{\rmfamily X}} where $X$
is the placeholder name.\footnote{The placeholders will need to be
  commented in order to not interfere with the semantics of the
  process modelling language~(\cf Javadoc in Java).  In \Prism, we
  therefore use \texttt{//}\yaptplparam{\emph{\rmfamily X}}.}
\Cref{tab:tpl-param} lists placeholders currently supported by \Yap.

\begin{table}[t]
  \caption{Placeholders recognised by \Yap and to be
    inserted into the process model for substitution
    \label{tab:tpl-param}}
  \small 
  \begin{tabularx}{\textwidth}{lX}
    \toprule
    \textbf{Placeholder} & \textbf{Description} 
    \\\midrule
    \yaptplparam{TYPES} & Inject global type declarations \\
    \yaptplparam{PREDICATES} & Inject global definitions \\ 
    \yaptplparam{CONTROLLER} & Inject controller module \\
    \yaptplparam{REWARDS} & Inject reward structures \\
    \yaptplparam{MODULEHOOK(m)} & Add data and command
    definitions to application module \texttt{m}
    \\\bottomrule
  \end{tabularx}
\end{table}

The output of \Yap, for this use case, is an \ac{MDP}, with its
non-deterministic choice representing the decision space of all actors
in the work cell.  We call the decision space of the safety
controller---as one of these actors---the \emph{design space}.  This
design space and the decision space of the other actors are used for
optimal controller synthesis.

With the model constructed according to the steps \mitem[1] to \mitem[9] in
\Cref{fig:yap-workflow}, we use \Yap to generate and inject the
controller into the \emph{process
  model}~(\Cref{sec:step1-process-model}):
\begin{verbatim}
  yapp -m model.yap -t target-template.xyz -o output/model.xyz \
       -f prism -d multi-event-concurrent --synthesise controller
\end{verbatim}
With the output format switch \texttt{-f prism}, \Yap creates three
artefacts in this step:
\begin{enumerate}
\item An \ac{MDP} in \Prism's \ac{pGCL} used as the \emph{design
    space} for \Prism's search for an \emph{optimal policy}---a
  \ac{DTMC}---representing the abstract controller (file
  \texttt{model.prism}).
\item A list of \ac{PCTL} properties to be verified of the design
  space by \Prism in step \mitem[11]~(\Cref{sec:step8-synth-dtmc},
  file \texttt{model.props}).
\item A list of \ac{PCTL} properties to be verified in step
  \mitem[12]~(\Cref{sec:policy-verif}) of any policy found by
  \Prism~(file \texttt{model\_pol.props}).
\end{enumerate}

The following \ac{pGCL} fragment, generated by \Yap, shows the design
space for \emph{switching into a safety mode} triggered if a particular
hazard~(\egs$\rfct[HC]$, $\rfct[HRW]$, $\rfct[HS]$) has been detected:
\begin{lstlisting}[language=prism]
[si_HCSrmstIdleVissafmod] !CYCLEEND & safmod=normal & HCp=act -> (safmod'=srmst);
[si_HCStOffAudsafmod] !CYCLEEND & safmod=normal & HCp=act -> (safmod'=stopped);
[si_HCStOffVissafmod] !CYCLEEND & safmod=normal & HCp=act -> (safmod'=stopped);
[si_HRWmitsafmod] !CYCLEEND & safmod=normal & HRWp=act -> (safmod'=srmst);
[si_srmstsafmod] !CYCLEEND & safmod=normal & HSp=act -> (safmod'=srmst);
\end{lstlisting}

The following \ac{pGCL} fragment, generated by \Yap, highlights the
part of the controller design space allowing the controller to \emph{switch
off mode-specific safety functions} (lines 1-2) and \emph{resume to a
less restrictive safety mode} (lines 3-4):
\begin{lstlisting}[language=prism,emph={CE\_HC,hST\_HOinSGA}]
[si_HCres2fun] !CYCLEEND & HCp=mit & !CE_HC & notif=leaveArea & !hST_HOinSGA -> (notif'=ok);
[si_HCresfun] !CYCLEEND & HCp=mit & !CE_HC & notif=leaveArea & !hST_HOinSGA -> (notif'=ok);
[si_HRWressafmod] !CYCLEEND & safmod=normal & HCp=inact & HSp=act & WSp=inact & HWp=inact & RCp=inact & (RTp=mit | RTp=sfd) & HRWp=mit & hloc!=sharedTbl & (notif_leaveWrkb=false) 
     -> (safmod'=ssmon)&(HRWp'=sfd);
\end{lstlisting}

\paragraph{\mitem[11] Design Space Verification and Optimal Controller
  Synthesis.}
\label{sec:step8-synth-dtmc}

In this example, we use \Prism for the verification of the \ac{MDP}
and the synthesis of an \ac{MDP} policy.
The use of \Prism and the processing of its output is out of scope of
this paper and, therefore, not explained here.  Particularly, the
formulas presented below contain \ac{PCTL}
operators~\cite{Kwiatkowska2007-StochasticModelChecking} and other
\Prism query language~\cite{Parker2019-PRISMModelChecker} primitives
that are assumed to be familiar to the \Yap user interested in
\Prism-based controller synthesis with \Yap.

\paragraph{Synthesising Optimal Policies from an \ac{MDP}.}

For optimal policy synthesis~(here, the synthesis of \acp{DTMC}), we
use the command
\begin{verbatim}
  prism output/model.prism -pctl '<query>' -s \
       -exportadvmdp poloutdir/model-adv.tra \
       -exportstates poloutdir/model-adv.sta \
       -exportprodstates poloutdir/model-adv.pst \
       -exportlabels poloutdir/model-adv.lab .
\end{verbatim}
\Prism will generate one or more policy files~(with names and suffixes
according to \texttt{model-adv[1-n]. {pst,sta,tra,lab}}) from the file
\texttt{output/model.prism} and compliant with the optimisation query
\texttt{<query>}, \eg
\begin{verbatim}
  multi(R{"effort"}max=? [ C ], R{"nuisance"}max=? [ C ]) ,
\end{verbatim}
and places these files in \texttt{poloutdir/}.  This query searches
for all policies that Pareto-maximise effort and nuisance
(\texttt{Rmax[C]}) as explained in \Cref{sec:verif-contr-synth}.
\Prism enumerates such policies as a list of value pairs holding the
results of the cost function defined by this query.  Within \Prism's
GUI, one can visualise these pairs as a Pareto front.  We calculated a
Pareto front for the present example in
\cite{Gleirscher2020-SafetyControllerSynthesis}.

To include the verification of safety properties at this
stage in the procedure, we can use a combination of a single
optimisation query and several constraints,
\eg
\begin{verbatim}
  multi(R{"prod"}max=? [ C ], R{"risk_sev"}<=s [ C ]) and
  multi(R{"risk_sev"}<=s [ C<=t ], P<=p [ F "ANY" ]) .
\end{verbatim}
The first property maximises the \texttt{prod}uctivity of the work
cell (\texttt{Rmax[C]}) as long as the accumulative bound on action
rewards (\texttt{R<=s[C]}) for \texttt{risk_sev} stays below a
user-defined level $s$.  The second property combines a time-bounded
version (\texttt{C<=t}) of the latter constraint with the
probability-bounded reachability (\texttt{P<=p[F]}) of \texttt{ANY} of
the modelled factors, for user-defined bounds $t$ and $p$.
Beyond \texttt{ANY}, \Yap generates further shortcut formulas (into
the file \texttt{model.prism}) that can be used in \ac{PCTL}
properties as shown above.  These state formulas are listed in
\Cref{tab:formulas}.

\begin{table}
  \caption{Formulas generated by \Yap for custom property specification
    \label{tab:formulas}}
  \small 
  \begin{tabularx}{\textwidth}{>{\ttfamily}lX}
    \toprule {\rmfamily\textbf{Formula}} & \textbf{Description}
    \\\midrule
    $E$ & ``detector'' predicate for the \acl{CE} $E$ \\
    RCE\_$E$ & ``ground truth or reality'' predicate for the \acl{CE} $E$ \\
    ANYOCC (OCE) & \emph{universal} detector predicate, true if
    \emph{any} critical event is true
    \\
    ANYREC (RCE) & universal ground truth counterpart of \texttt{ANYOCC} \\
    ANY (CE) & true if any \ac{CE} has occurred whether
    or not detected \\
    ACCIDENT & true if any factor $\rfct$ is in its mishap phase
    $\mishp$ \\
    MISHAP & true if \texttt{ACCIDENT} or if any \yapkeyw{final}
    factor (\egs an incident) is activated (\ies in phase $\activ$) \\
    SAFE & true of any state that is neither a \ac{CE} nor a mishap \\
    FINAL (CYCLEEND) & true if
    \texttt{hFINAL\_CUSTOM}~(\Cref{sec:step3-mode-specs}) is reached
    \\\bottomrule
  \end{tabularx}
\end{table}

\paragraph{\mitem[12] Controller Verification: Checking the Generated
  Policies (\acp{DTMC}).}
\label{sec:policy-verif}

As already mentioned in \Cref{sec:overview}, the separation into the
two verification steps \mitem[11] and \mitem[12] and the corresponding
property lists is due to restrictions in the combinations of
properties that can be checked by \Prism in one go.  
The policies are given in the form of \acp{DTMC} and can, thus, be
further checked with the command
\begin{verbatim}
  prism -importstates poloutdir/model-adv.sta \
        -importlabels poloutdir/model-adv.lab \
        -importtrans poloutdir/model-adv1.tra -dtmc \
        -pctl "<prop>" -gs >poloutdir/model-adv-checks.txt .
\end{verbatim}
\Eg for \texttt{<prop>} we
applied
\begin{verbatim}
  filter(avg, P=? [ !"ACCIDENT" W "SAFE" ], "ANYREC" & !"MISHAP")
\end{verbatim}
to determine the average probability~(\texttt{filter(avg, P=? [...],
  ...)}) of accident freedom until reaching a safe
state~(\verb+!"ACCIDENT" W "SAFE"+) when starting from any reachable
hazardous state (\verb+"ANYREC" \& !"MISHAP"+).  As already mentioned
in \Cref{sec:step7-synth-mdp}, with the file \texttt{model.props},
\Yap suggests a range of properties to be checked of a
policy.  See \cite[Tab.~III]{Gleirscher2020-SafetyControllerSynthesis}
for a selection of properties.

\paragraph{Further Processing of the \ac{MDP} and \ac{DTMC}.}
\label{sec:ctr-ref}

The abstract controller consists of the list of states of the
\ac{MDP} respectively the \ac{DTMC}, \eg\\
\texttt{State:(... HRWp, \textbf{notif_leaveWrkb}, ...)}\\
\texttt{510:   (... 1, \textbf{false}, ...)}\\
\texttt{511:   (... 1, \textbf{true}, ...)}\\
and all transitions describing the decisions (accordingly, with probabilistic outcomes), \eg
\begin{verbatim}
510 511 1 si_HRWmitfun
\end{verbatim}
This transition, going from state 510 to 511 and labelled with the
action \texttt{si\_HRWmitfun}, notifies the operator, with probability
1,\footnote{The controller in this example is fully deterministic but,
  conceptually, we can also design randomised controllers with our
  approach.} to leave the workbench in case of an activated factor
$\rfct[HRW]$.

For visualisation and model debugging, the \ac{MDP}
can be converted into a \texttt{dot} file with
\begin{verbatim}
  prism output/model.prism -exporttransdotstates rel.dot .
\end{verbatim}
This file can be used with GraphViz tools such as
\texttt{dot}.\footnote{See \url{https://graphviz.org}.}  Based on
GraphViz, \Yap provides rudimentary facilities for the visualisation
of the generated policy.  Such a visualisation can be useful in
the direct debugging of \acp{DTMC} with a state space of up
to a size of around 1000 states.

\section{Discussion and Outlook}
\label{sec:outlook}

\paragraph{From Abstract to Concrete Policies.}

The safety controller in its abstract form is represented by the
calculated policy, a \ac{DTMC} with state space $\Sigma$ and action
set $A$.  $\Sigma$ and $A$ are results of combining the risk state
space, generated by \Yap from the factor set $F$, and the mitigation
actions, filed in the \Yap model, with the process model.  Each
state-based memory-less deterministic policy $\pi$ can then be
represented by a map $\pi\colon\Sigma\to A$.  As shown before, the
transition relation of the policy is provided by \Prism as a list of
$(\mathit{state, action, probability,
  state})$-tuples~(\cf\Cref{sec:ctr-ref}).  The safety controller, part
of such a policy, is a list of transitions that, at the concrete level,
would again be guarded commands of the form:
\begin{align*}
  \underbrace{[\mbox{controller action}]}_{\text{event}}\;
  \underbrace{\mbox{process \& risk state}}_{\text{guard}}
  & \rightarrow
    \underbrace{\text{mode \& activity switch, safety function}}_{\text{update}}
\end{align*}

\Prism's output can be used to translate this abstract policy
representing the discrete-event controller into a concrete policy.
This translation involves two essential steps:
\begin{itemize}
\item The translation of the abstract states into concrete guard
  conditions, and
\item the translation of the updates into low-level procedures
  generating control inputs to the process.
\end{itemize}
Part of ongoing research is the corresponding refinement of this
transition relation into an automaton that can run on, \eg an
autonomous machine platform or the \ac{ROS}.  We will investigate how
environments such as
Isabelle/UTP~\cite{Foster2015-IsabelleUTPMechanisedTheory} and
\textsc{RoboTool}~\cite{Miyazawa2019-RoboChartmodellingverification}
can be used to verify and deploy safety controllers derived with the
help of \Yap.  Isabelle/UTP provides a generic framework for model
verification and \textsc{RoboTool} an environment for rigorous robotic
software development.

\paragraph{Safety Properties and Safety Controller.}

A safety property states that ``something will \emph{not} happen''
and, thus, is a property whose violation can be observed in finite
time~\cite{Lamport1977-ProvingCorrectnessMultiprocess}.  In many
applications, ``something'' refers to a \ac{CE} that cannot be avoided
by a careful redesign of the process and, therefore, violations have to
be accepted to a certain extent.  A safety controller typically
includes a safety monitor responsible for detecting such violations at
run-time~\cite{Leucker2009-briefaccountruntime} and an active
component influencing the monitored process in a way that the safety
property is established again.  In other words, the ``violation counter''
is restored.  In such applications, we therefore substitute the
verification of the original safety property by
\begin{itemize}
\item a response
  property~\cite{Manna1995-TemporalVerificationReactive}~(formalising
  successful mitigation and resumption as a finite response to the
  detection of an endangerment) to be verified of the process
  integrated with the controller design space~(\cf\mitem[11]), and
\item another safety property~(\cf\mitem[12], formalising the absence
  of undesired consequences of the aforementioned violations) whose
  probability of being violated must not exceed a certain bound, by
  virtue of the safety controller when working correctly.
\end{itemize}

\paragraph{Re-Interpretation of Activity Graphs for Synthesis.}

We may want to allow several actors to concurrently carry out actions
in any of the activities of the process.  Therefore, it seems useful
to associate a \ac{CPN}~\cite{Jensen2009-ColouredPetriNets} semantics
to activity graphs~(\egs \Cref{fig:exa-hrc-off}).  \acp{CPN} offer a
more flexible way of modelling concurrency compared to the parallel
composition~\cite{Hoare1985-CommunicatingSequentialProcesses} used in
\Prism's \ac{pGCL}.  Specifically, in a \ac{CPN}, the places could
represent activities and the movable labels the actors.  Then, a
placement of these labels, \ie a marking, indicates the activities
actors are performing at a point in time.  A transition in a \ac{CPN}
can move any number of labels between the activities, meaning the
actors involved in that transition concurrently finish their current
activities and start new activities.  However, a \ac{pGCL} guarded
command in \Prism can either move one label or as many labels as there
are actors participating in a synchronous event.  As a part of our
future work, we will investigate how the explicit approach to
concurrency in \acp{CPN} improves the usefulness and flexibility of
the activity model.

\section{Conclusions}
\label{sec:conclusions}

This paper provides a hands-on and tool-focused guide to a novel
approach to the design, verification, and synthesis of safety
controllers from hazard analysis and risk assessment as previously
published in \cite{Gleirscher2020-SafetyControllerSynthesis}.  We also
discuss a range of modelling decisions~(\egs identifying parameters,
decomposing behaviour, integrating probabilistic choice) to be made
when devising such controllers.  The proposed step-wise and
tool-supported workflow aims at supporting verification engineers in
transforming data from \acl{HARA} into a verifiable controller model
and, thus, contributes to recent and practically relevant
challenges~(\egs\cite[Challenges OC1, OC2, and
OC4]{Calinescu2019-SocioCyberPhysical}).

Among the next steps of technical research are the improvement of the
synthesis facilities, the evaluation of alternatives to \Prism, and
the development of an integration with robotic platforms~(\egs
\ac{ROS}, digital twin environments\footnote{See, \egs
  \url{https://github.com/douthwja01/CSI-cobotics}.}) and tools~(\egs
\textsc{RoboChart}\cite{Miyazawa2019-RoboChartmodellingverification})
to automate controller deployment.

\paragraph*{Acknowledgements.}

This research was funded by the Lloyd's Register Foundation under the
Assuring Autonomy International Programme grant CSI:Cobot.  I am
greatly indebted to Radu Calinescu for many inspiring discussions and
for encouraging me to implement some of the described \Yap enhancements.

\nocite{*}
\bibliographystyle{eptcs}
\bibliography{}
\end{document}